# A Hardware-oriented Algorithm for Complex-valued Constant Matrix-vector Multiplication

Aleksandr CARIOW[1], Galina CARIOWA[1]

[1]West Pomeranian University of Technology, Szczecin, 720229, Poland

atariov@wi.zut.edu.pl

*Abstract*— In this communication we present a hardware-oriented algorithm for constant matrix-vector product calculating, when the all elements of vector and matrix are complex numbers. The main idea behind our algorithm is to combine the advantages of Winograd's inner product formula with Gauss's trick for complex number multiplication. The proposed algorithm versus the naïve method of analogous calculations drastically reduces the number of multipliers required for FPGA implementation of complex-valued constant matrix-vector multiplication. If the fully parallel hardware implementation of naïve (schoolbook) method for complex-valued matrix-vector multiplication requires $4MN$ multipliers, $2M$ $N$-inputs adders and $2MN$ two-input adders, the proposed algorithm requires only $3N(M+1)/2$ multipliers and $[3M(N+2)+1,5N+2]$ two-input adders and $3(M+1)$ $N/2$-input adders.

*Index Terms*—algorithm design and analysis, signal processing algorithms, digital signal processing chips, high performance computing.

## I. INTRODUCTION

Most of the computation algorithms which are used in digital signal, image and video processing, computer graphics and vision and high performance supercomputing applications have matrix-vector multiplication as the kernel operation [1, 2]. For this reason, the rationalization of these operations is devoted to numerous publications [3-18]. In some cases, elements of the multiplied matrices and vectors are complex numbers [5-9]. In the general case a fully parallel hardware implementation of a rectangular complex-valued matrix-vector multiplication requires $MN$ multipliers of complex numbers. In the case where the matrix elements are constants, we can use encoders instead of multipliers. This solution greatly simplifies implementation, reduces the power dissipation and lowers the price of the device. On the other hand, when we are dealing with FPGA chips that contain several tens or even hundreds of embedded multipliers, the building and using of additional encoders instead of multipliers is irrational. Examples could be that of the Xilinx Spartan-3 family of FPGA's which includes between 4 and 104 18x18 on-chip multipliers and the Altera Cyclone-III family of FPGA's which include between 23 and 396 18×8 on-chip multipliers. Another Altera's Stratix-V GS family of FPGA's has between 600 and 1963 variable precision on-chip blocks optimized for 27×27 bit multiplication. In this case, it would be unreasonable to refuse the possibility of using embedded multipliers. Nevertheless, the number of on-chip multipliers is always limited, and this number may sometimes not be enough to implement a high-speed fully parallel matrix-vector multiplier. Therefore, finding ways to reduce the number of multipliers in the implementation of matrix-vector multiplier is an extremely urgent task. Some interesting solutions related to the rationalization of the complex-valued matrix-matrix and matrix-vector multiplications have already been obtained [10-13]. There are also original and effective algorithms for constant matrix-vector multiplication. However, the rationalized algorithm for complex-valued constant matrix-vector multiplications has not yet been published. For this reason, in this paper, we propose such algorithm.

## II. PRELIMINARY REMARKS

The complex-valued vector-matrix product may be defined as:

$$\mathbf{Y}_{M \times 1} = \mathbf{A}_{M \times N} \mathbf{X}_{N \times 1} \quad (1)$$

where $\mathbf{X}_{N \times 1} = [x_0, x_1, ..., x_{N-1}]^T$ - is $N$-dimensional complex-valued input vector, $\mathbf{Y}_{N \times 1} = [y_0, y_1, ..., y_{M-1}]^T$ - is $N$-dimensional complex-valued output vector, and

$$\mathbf{A}_{M \times N} = \begin{bmatrix} a_{0,0} & a_{0,1} & \cdots & a_{0,N-1} \\ a_{1,0} & a_{1,1} & \cdots & a_{1,N-1} \\ \vdots & \vdots & \ddots & \vdots \\ a_{M-1,0} & a_{M-1,1} & \cdots & a_{M-1,N-1} \end{bmatrix},$$

where $n = 0, 1, ..., N-1$, $m = 0, 1, ..., M-1$, and $x_n = x_n^{(r)} + jx_n^{(i)}$, $a_{m,n} = a_{m,n}^{(r)} + ja_{m,n}^{(i)}$, $y_m = y_m^{(r)} + jy_m^{(i)}$.

In this expression $x_n^{(r)}$, $x_n^{(i)}$, $y_m^{(r)}$, $y_m^{(i)}$ are real variables, $a_{m,n}^{(r)}$, $a_{m,n}^{(i)}$ are real constants, and $j$ is the imaginary unit, satisfying $j^2 = -1$. Superscript $r$ means the real part of complex number, and the superscript $i$ means the imaginary part of complex number. The task is to calculate the product defined by the expression (1) with the minimal multiplicative complexity.

## III. BRIEF BACKGROUND

It is well known, that complex multiplication requires four real multiplications and two real additions, because:

$$(a+jb)(c+jd) = ac - bd + j(ad + bc) \quad (2)$$

So, we can observe that the direct computation of (1) requires $NM$ complex multiplications ($4NM$ real multiplications) and $2M(2N-1)$ real additions.

According to Winograd's formula for inner product calculation each element of vector $\mathbf{Y}_{M \times 1}$ can be calculated as follows [16]:

$$y_m = \sum_{k=0}^{\frac{N}{2}-1}[(a_{m,2k}+x_{2k+1})(a_{m,2k+1}+x_{2k})] - c_m - \xi_M \quad (3)$$

where

$$c_m = \sum_{k=0}^{\frac{N}{2}-1} a_{m,2k} \cdot a_{m,2k+1} \text{ and } \xi_N = \sum_{k=0}^{\frac{N}{2}-1} x_{2k} \cdot x_{2k+1}$$

if $N$ is even. (The case of odd $N$, we will not be considered here, as it can easily be reduced to the even length $N$). It is clear that if we are dealing with complex-valued data, then $c_m = c_m^{(r)} + jc_m^{(i)}$ and $\xi_N = \xi_N^{(r)} + j\xi_N^{(i)}$, where are $\xi_N^{(r)}$ and $\xi_N^{(i)}$ are real and imaginary parts of calculated real variable $\xi_N$ respectively, $c_m^{(r)}$ and $c_m^{(i)}$ are real and imaginary parts of calculated in advance constants $c_m$. Here it should be emphasized that because $a_{m,n}$ are constants, the $c_m$ can be precomputed and stored in a lookup table in advance. Thus, the calculation of $c_m$ does not require the execution of arithmetic operations during realization of the algorithm. The calculation of $\xi(N)$ requires the implementation of the $N/2$ complex multiplications. Therefore, we can observe that the computation of (3) for all $m$ requires only $N(M+1)/2$ complex multiplications ($2N(M+1)$ real multiplications). However, the number of real additions in this case is significantly increased.

It is well known too, that the complex multiplication can be carried out using only three real multiplications and five real additions, because [13]:

$$(a+jb)(c+jd) = ac - bd + j[(a+b)(c+d) - ac - bd] \quad (4)$$

Expression (4) is well known as Gauss's trick for multiplication of complex numbers [17]. Taking into account this trick the expression (3) can be calculated using the only $3N(M+1)/2$ multiplications of real numbers at the expense of further increase in the number of real additions.

## IV. THE ALGORITHM

First, we present the vector $\mathbf{X}_{N\times 1} = [x_0, x_1, ..., x_{N-1}]^T$ in a following form: $\mathbf{X}_{2N\times 1} = [x_0^{(r)}, x_0^{(i)}, x_1^{(r)}, x_1^{(i)}, ..., x_{N-1}^{(r)}, x_{N-1}^{(i)}]^T$, and vector $\mathbf{Y}_{M\times 1} = [y_0, y_1, ..., y_{M-1}]$ - in a following form: $\mathbf{Y}_{2M\times 1} = [y_0^{(r)}, y_0^{(i)}, y_1^{(r)}, y_1^{(i)}, ..., y_{N-1}^{(r)}, y_{N-1}^{(i)}]^T$.

Next, we splits vector $\mathbf{X}_{2N\times 1}$ into two vectors $\mathbf{X}_{N\times 1}^{(1)}$ and $\mathbf{X}_{N\times 1}^{(2)}$ containing only even-numbered and only odd-numbered elements respectively:

$$\mathbf{X}_{N\times 1}^{(1)} = [x_0^{(r)}, x_0^{(i)}, x_2^{(r)}, x_2^{(i)}, ..., x_{N-2}^{(r)}, x_{N-2}^{(i)}]^T,$$
$$\mathbf{X}_{N\times 1}^{(2)} = [x_1^{(r)}, x_1^{(i)}, x_3^{(r)}, x_3^{(i)}, ..., x_{N-1}^{(r)}, x_{N-1}^{(i)}]^T.$$

Then from the elements of the matrix . we form two super-vectors of data:

$$\mathbf{A}_{MN\times 1}^{(1)} = [\hat{\mathbf{A}}_{2M\times 1}^{(0)}, \hat{\mathbf{A}}_{2M\times 1}^{(1)}, ..., \hat{\mathbf{A}}_{2M\times 1}^{(\frac{N}{2}-1)}]^T,$$

$$\mathbf{A}_{MN\times 1}^{(2)} = [\breve{\mathbf{A}}_{2M\times 1}^{(0)}, \breve{\mathbf{A}}_{2M\times 1}^{(1)}, ..., \breve{\mathbf{A}}_{2M\times 1}^{(\frac{N}{2}-1)}]^T,$$

where

$$\hat{\mathbf{A}}_{2M\times 1}^{(k)} = [a_{0,2k+1}^{(r)}, a_{0,2k+1}^{(i)}, a_{1,2k+1}^{(r)}, a_{1,2k+1}^{(i)}, ..., a_{M-1,2k+1}^{(r)}, a_{M-1,2k+1}^{(i)}]^T,$$

$$\breve{\mathbf{A}}_{2M\times 1}^{(k)} = [a_{0,2k}^{(r)}, a_{0,2k}^{(i)}, a_{1,2k}^{(r)}, a_{1,2k}^{(i)}, ..., a_{M-1,2k}^{(r)}, a_{M-1,2k}^{(i)}]^T,$$

And now we introduce the vectors

$$\mathbf{C}_{2M\times 1} = [c_0^{(r)}, c_0^{(i)}, c_1^{(r)}, c_1^{(i)}, ..., c_{M-1}^{(r)}, c_{M-1}^{(i)}]^T,$$

$$\mathbf{\Xi}_{2M\times 1} = [\xi_N^{(r)}, \xi_N^{(i)}, \xi_N^{(r)}, \xi_N^{(i)}, ..., \xi_N^{(r)}, x_N^{(i)}]^T.$$

Next, we introduce some auxiliary matrices:

$$\mathbf{P}_{MN\times N} = \mathbf{I}_{\frac{N}{2}} \otimes (\mathbf{1}_{M\times 1} \otimes \mathbf{I}_2), \quad \breve{\mathbf{T}}_{\frac{3}{2}MN\times MN} = \mathbf{I}_{\frac{MN}{2}} \otimes \mathbf{T}_{3\times 2},$$

$$\mathbf{\Sigma}_{3M\times \frac{3MN}{2}} = \mathbf{1}_{1\times \frac{N}{2}} \otimes (\mathbf{I}_M \otimes \mathbf{I}_3), \quad \hat{\mathbf{T}}_{2M\times 3M} = \mathbf{I}_M \otimes \mathbf{T}_{2\times 3},$$

$$\mathbf{T}_{3\times 2} = \begin{bmatrix} 1 & \\ & 1 \\ 1 & -1 \end{bmatrix}, \quad \mathbf{T}_{2\times 3} = \begin{bmatrix} 1 & & 1 \\ & 1 & 1 \end{bmatrix}.$$

where $\mathbf{1}_{M\times N}$ - is an $M\times N$ matrix of ones (a matrix where every element is equal to one), $\mathbf{I}_N$ - is an identity $N\times N$ matrix and sign „$\otimes$" denotes tensor product of two matrices [18].

Using the above matrices the rationalized computational procedure for calculating the constant matrix-vector product can be written as follows:

$$\mathbf{Y}_{2M\times 1} = \mathbf{\Xi}_{2M\times 1} + \{\mathbf{C}_{2M\times 1} + \hat{\mathbf{T}}_{2M\times 3M}[\mathbf{\Sigma}_{3M\times \frac{3}{2}MN} \times \mathbf{D}_{\frac{3}{2}MN} \breve{\mathbf{T}}_{\frac{3}{2}MN\times MN} (\mathbf{A}_{MN\times 1}^{(1)} + \mathbf{P}_{MN\times N} \mathbf{X}_{N\times 1}^{(1)})]\} \quad (5)$$

$$\mathbf{D}_{\frac{3}{2}MN} = \bigoplus_{l=0}^{\frac{MN}{2}-1} \mathbf{D}_3^{(l)}, \quad \mathbf{D}_3^{(l)} = diag(s_0^{(l)}, s_1^{(l)}, s_2^{(l)}),$$

where sign „$\oplus$" denotes direct sum of the matrices which are numbered in accordance with the increase of the superscript value [18].

If the elements of $\mathbf{D}_{\frac{3}{2}MN}$ placed vertically without disturbing the order and written in the form of the vector $\mathbf{S}_{\frac{3}{2}MN\times 1} = \mathbf{D}_{\frac{3}{2}MN} \mathbf{1}_{\frac{3}{2}MN\times 1}$, then they can be calculated using the following vector-matrix procedure:

$$\mathbf{S}_{\frac{3}{2}MN\times 1} = \tilde{\mathbf{T}}_{\frac{3}{2}MN\times MN} (\mathbf{A}_{MN\times 1}^{(2)} + \mathbf{P}_{MN\times N} \mathbf{X}_{N\times 1}^{(2)}) \quad (6)$$

$$\tilde{\mathbf{T}}_{\frac{3}{2}MN\times MN} = \mathbf{I}_{\frac{MN}{2}} \otimes \tilde{\mathbf{T}}_{3\times 2}, \quad \mathbf{T}_{3\times 2} = \begin{bmatrix} 1 & -1 \\ 1 & 1 \\ 0 & 1 \end{bmatrix}.$$

As already noted, the elements of the vector $\mathbf{C}_{2M\times 1}$ can be calculated in advance. However, the elements of vector $\mathbf{\Xi}_{2M\times 1}$ must be calculated during the realization of the algorithm. The procedure describes the implementation of computing elements of this vector can be represented in the following form:

$$\mathbf{\Xi}_{2M\times 1} = \mathbf{P}_{2M\times 2} \mathbf{T}_{2\times 3} \mathbf{\Sigma}_{3\times \frac{3N}{2}} \mathbf{\Psi}_{\frac{3N}{2}} \breve{\mathbf{T}}_{\frac{3N}{2}\times N} \mathbf{X}_{N\times 1}^{(1)} \quad (7)$$

where

$$\breve{\mathbf{T}}_{\frac{3N}{2}\times N} = \mathbf{I}_{\frac{N}{2}} \otimes \mathbf{T}_{3\times 2}, \quad \mathbf{P}_{2M\times 2} = \mathbf{1}_{M\times 1} \otimes \mathbf{I}_2, \quad \mathbf{\Sigma}_{3\times \frac{3N}{2}} = \mathbf{1}_{1\times \frac{N}{2}} \otimes \mathbf{I}_3,$$

and $\Psi_{\frac{3}{2}N} = \bigoplus_{k=0}^{\frac{N}{2}-1} \Psi_3^{(k)}$, $\Psi_3^{(k)} = diag(\varepsilon_0^{(2k+1)}, \varepsilon_1^{(2k+1)}, \varepsilon_2^{(2k+1)})$.

If the elements of $\Psi_{\frac{3}{2}N}$ placed vertically without disturbing the order and written in the form of the vector $E_{\frac{3}{2}N \times 1} = \Psi_{\frac{3}{2}N} 1_{\frac{3}{2}N \times 1}$, then they can be calculated using the following vector-matrix procedure:

$$E_{\frac{3}{2}N \times 1} = \tilde{T}_{\frac{3}{2}N \times N} X_{N \times 1}^{(2)}), \ \tilde{T}_{\frac{3}{2}N \times N} = I_{\frac{N}{2}} \otimes \tilde{T}_{3 \times 2}.$$

Consider, for example, the case of $N = 4$ and $M = 3$. Then the procedure (5) takes the following form:

$$Y_{6 \times 1} = \Xi_{6 \times 1} + \{C_{6 \times 1} + [\Sigma_{9 \times 18} \times D_{18} T_{18 \times 12} (A_{12 \times 1}^{(1)} + P_{12 \times 4} X_{4 \times 1}^{(1)})]\},$$

where

$Y_{6 \times 1} = [y_0^{(r)}, y_0^{(i)}, y_1^{(r)}, y_1^{(i)}, y_3^{(r)}, y_3^{(i)}]^T$,

$X_{4 \times 1}^{(1)} = [x_0^{(r)}, x_0^{(i)}, x_2^{(r)}, x_2^{(i)}]^T$, $X_{4 \times 1}^{(2)} = [x_1^{(r)}, x_1^{(i)}, x_3^{(r)}]^T$,

$D_{18} = \bigoplus_{l=0}^{5} D_3^{(l)}$, $D_3^{(l)} = diag(s_0^{(l)}, s_1^{(l)}, s_2^{(l)})$,

$S_{18 \times 1} = D_{18} 1_{18 \times 1} = \tilde{T}_{18 \times 12} (A_{12 \times 1}^{(2)} + P_{12 \times 4} X_{4 \times 1}^{(2)})$,

$A_{12 \times 1}^{(1)} = [\hat{A}_{6 \times 1}^{(0)}, \hat{A}_{6 \times 1}^{(1)}]^T$, $A_{12 \times 1}^{(2)} = [\check{A}_{6 \times 1}^{(0)}, \check{A}_{6 \times 1}^{(1)}]^T$

$\Xi_{6 \times 1} = [\xi_4^{(r)}, \xi_4^{(i)}, \xi_4^{(r)}, \xi_4^{(i)}, \xi_4^{(r)}, \xi_4^{(i)}]^T$, $\Sigma_{9 \times 18} = 1_{1 \times 2} \otimes I_9$

$\hat{A}_{6 \times 1}^{(0)} = [a_{0,1}^{(r)}, a_{0,1}^{(i)}, a_{1,1}^{(r)}, a_{1,1}^{(i)}, a_{2,1}^{(r)}, a_{2,1}^{(i)}]^T$, $\tilde{T}_{18 \times 12} = I_6 \otimes \tilde{T}_{3 \times 2}$

$\hat{A}_{6 \times 1}^{(1)} = [a_{0,3}^{(r)}, a_{0,3}^{(i)}, a_{1,3}^{(r)}, a_{1,3}^{(i)}, a_{2,3}^{(r)}, a_{2,3}^{(i)}]^T$, $\tilde{T}_{18 \times 12} = I_6 \otimes T_{3 \times 2}$,

$\check{A}_{6 \times 1}^{(0)} = [a_{0,0}^{(r)}, a_{0,0}^{(i)}, a_{1,0}^{(r)}, a_{1,0}^{(i)}, a_{2,0}^{(r)}, a_{2,0}^{(i)}]^T$, $\hat{T}_{6 \times 9} = I_3 \otimes T_{2 \times 3}$,

$\check{A}_{6 \times 1}^{(1)} = [a_{0,2}^{(r)}, a_{0,2}^{(i)}, a_{1,2}^{(r)}, a_{1,2}^{(i)}, a_{2,2}^{(r)}, a_{2,2}^{(i)}]^T$, $P_{12 \times 4} = I_2 \otimes (1_{3 \times 1} \otimes I_2)$,

$C_{6 \times 1} = [c_0^{(r)}, c_0^{(i)}, c_1^{(r)}, c_1^{(i)}, c_2^{(r)}, c_2^{(i)}]^T$, $\tilde{T}_{6 \times 4} = I_2 \otimes \tilde{T}_{3 \times 2}$

$\Xi_{6 \times 1} = P_{6 \times 2} T_{2 \times 3} \Sigma_{3 \times 6} \Psi_6 \check{T}_{6 \times 4} X_{4 \times 1}^{(1)}$, $E_{6 \times 1} = \Psi_6 1_{6 \times 1} = \tilde{T}_{6 \times 4} X_{4 \times 1}^{(2)}$,

$\check{T}_{6 \times 4} = I_2 \otimes T_{3 \times 2}$, $P_{6 \times 2} = 1_{3 \times 1} \otimes I_2$, $\Sigma_{3 \times 6} = 1_{1 \times 2} \otimes I_3$,

and $\Psi_6 = \bigoplus_{k=0}^{1} \Psi_3^{(k)}$, $\Psi_3^{(k)} = diag(\varepsilon_0^{(2k+1)}, \varepsilon_1^{(2k+1)}, \varepsilon_2^{(2k+1)})$,

The data flow diagram for realization of proposed algorithm is illustrated in Figure 1. In turn, Figure 2 shows a data flow diagram for computing elements of the matrix $D_{3MN/2}$ in accordance with the procedure (6). In this paper, the data flow diagrams are oriented from left to right. Note [13-15] that the circles in these figures show the operation of multiplication by a real number (variable) inscribed inside a circle. Rectangles denote the real additions with values inscribed inside a rectangle. Straight lines in the figures denote the operation of data transfer. At points where lines converge, the data are summarized. (The dashed lines indicate the subtraction operation). We use the usual lines without arrows specifically so as not to clutter the picture. Figure 3a shows a data flow diagram for computing elements of the vector $\Xi_{2M \times 1}$ in accordance with the procedure (7). In turn, Figure 3b shows a data flow diagram for computing elements of the diagonal matrix $\Psi_{3N/2}$.

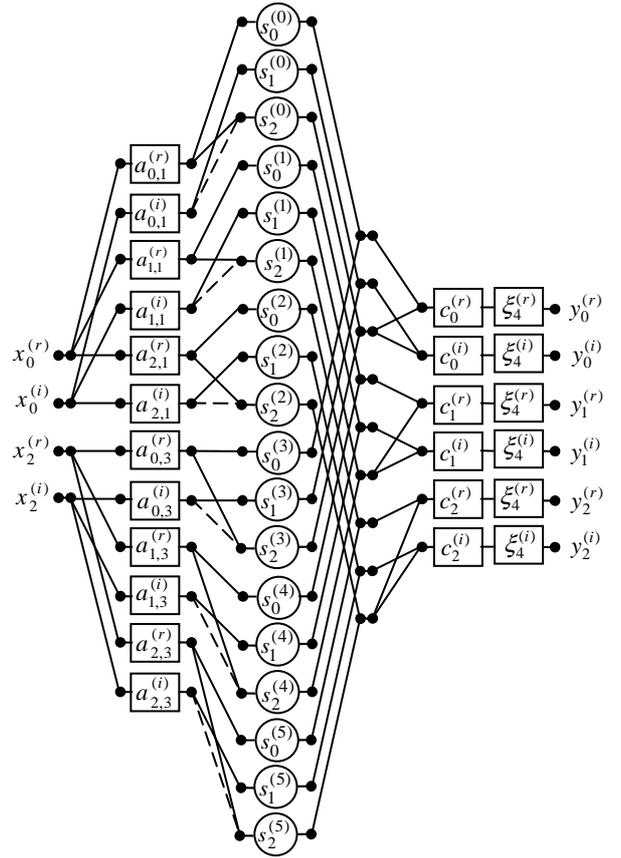

Figure 1. Data flow diagram for rationalized complex-valued constant matrix-vector multiplication algorithm for $N=4$, $M=3$.

## V. DISCUSSION OF HARDWARE COMPLEXITY

We calculate how many multipliers and adders are required, and compare this with the number required for a fully parallel naïve implementation of complex-valued matrix–vector product in Eq. (1). The number of conventional two-input multipliers required using the proposed algorithm is $3N(M+1)/2$. Thus using the proposed algorithm the number of multipliers to implement the complex-valued constant matrix-vector product is drastically reduced. Additionally our algorithm requires $2M(N+1)$ one-input adders with constant numbers (ordinary encoders), $M(N+4)+1,5N+2$ conventional two-input adders, and $3(M+1)$ $(N/2)$-input adders. Instead of encoders we can apply the ordinary two-input adders. Then the implementation of the algorithm will requires $3N(M+1)/2$ multipliers $3M(N+2)+1,5N+2$ two-input signed adders and $3(M+1)$ $(N/2)$-input adders.

In turn, the number of conventional two-input multipliers required using fully parallel implementation of "schoolbook" method for complex-valued matrix-vector multiplication is $4MN$. This implementation also requires the $2M$ $N$-inputs adders and $2MN$ two-input adders. Thus, our proposed algorithm saves 50 and even more percent of two-input embedded multipliers but it significantly increases number adders compared with direct method of fully-parallel implementation. For applications where the "cost" of a multiplication is greater than that of an addition, the new algorithm is always more computationally efficient than direct evaluation of the matrix-vector product. This allows concluding that the suggested solution may be useful in a number of cases and have practical application allowing to minimize complex-valued constant matrix-vector multiplier's hardware implementation costs.

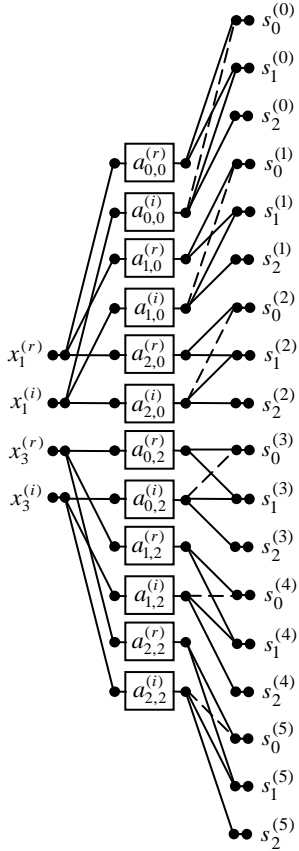

Figure 2. The data flow diagram for calculating elements of diagonal matrix $\mathbf{D}_{3MN/2}$ for $N=4$, $M=3$.

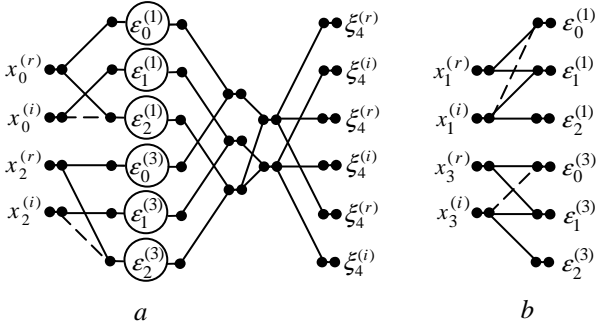

Figure 3. The data flow diagrams for calculating elements of vector $\mathbf{\Xi}_{6\times 1}$ (*a*), and for calculating elements of diagonal matrix $\mathbf{\Psi}_6$ (*b*).

## VI. Concluding Remarks

The article presents a new hardware-oriented algorithm for computing the complex-valued constant matrix-vector multiplication. To reduce the hardware complexity (number of two-operand multipliers), we exploit the Winograd's inner product formula and Gauss trick for complex number multiplication. This allows the effective use of parallelization of computations on the one hand and results in a reduction in hardware implementation cost of complex-valued constant matrix-vector multiplier on the other hand.

If the FPGA-chip already contains embedded multipliers, their number is always limited. This means that if the implemented algorithm contains a large number of multiplications, the developed processor may not always fit into the chip. So, the implementation of proposed in this paper algorithm on the base of FPGA chips, that have built-in binary multipliers, also allows saving the number of these blocks or realizing the whole complex-valued constant matrix-vector multiplying unit with the use of a smaller number of simpler and cheaper FGPA chips. It will enable to design of data processing units using a chips which contain a minimum required number of embedded multipliers and thereby consume and dissipate least power. How to implement a fully parallel complex-valued constant matrix-vector multiplier on the base of concrete FPGA platform is beyond the scope of this article, but it's a subject for follow-up articles.